\documentclass[a4paper,12pt]{article}
\usepackage{epsfig}
\usepackage{float}
\usepackage{graphicx}
\usepackage{amssymb}
\begin{document}
\thispagestyle{empty}

\newcommand{\etal}  {{\it{et al.}}}  
\def\Journal#1#2#3#4{{#1} {\bf #2}, #3 (#4)}
\def\PRD{Phys.\ Rev.\ D}
\def\NIMA{Nucl.\ Instrum.\ Methods A}
\def\PRL{Phys.\ Rev.\ Lett.\ }
\def\PLB{Phys.\ Lett.\ B}
\def\EPJ{Eur.\ Phys.\ J}
\def\IEEETNS{IEEE Trans.\ Nucl.\ Sci.\ }
\def\CPCD{Comput.\ Phys.\ Commun.\ }



{\Large\bf
\begin{center}
Dark photons in the decay of Higgs-like boson
\end{center}
}

\begin{center}
\large{ G.A. Kozlov  }
\end{center}
\begin{center}
 { Bogolyubov Laboratory of Theoretical Physics\\
 Joint Institute for Nuclear Research,\\
 Joliot Curie st., 6, Dubna, Moscow region, 141980 Russia }
\end{center}


 \begin{abstract}
 \noindent
 {We study the dark matter particle, the dark photon (DP), in the decay of the Higgs-like boson. The nature of dark matter is maintained through  the hidden sector  including the effects of breaking of the scale invariance. The model  is based on the additional $U^{\prime}(1)$ gauge group associated with light  DP. 
The interaction between DP and quarks is mediated by the derivative of the scalar - the dilaton. The latter appears in the conformal sector which triggers the electroweak symmetry breaking.
Upper limits are set on the DP mass, the mixing strength between the standard photon and DP. The model does allow to estimate the DP mass with the value of 4.5 MeV. The maximal value of the scale invariance breaking constant is also reported.  }


\end {abstract}




\bigskip

{\bf I. Introduction.-}  It is known the interest to  the theories in which the Standard Model (SM) is strongly coupled to a conformal sector of particles and the unparticle stuff. The certain gauge models may admit the additional $U^{\prime}(1)$ gauge group associated with new  gauge bosons  which can have the small masses or will be almost massless (for a recent review see [1] and the references therein).
The extended group $SU(2)_{L}\times U(1)_{Y}\times U^{\prime}(1)_{B}$ may appear, where  index $B$ in the new group $U^{\prime} (1)_{B}$ is associated with an extra gauge boson $B_{\mu}$ that would be the dark matter (DM), e.g., the dark photon (DP) $\gamma^{\star}$. The standard photon $\gamma$ may oscillate into  $\gamma^\star$ followed by the invisible decay to neutrino-antineutrino pair, $\gamma^{\star}\rightarrow \bar\nu\,\nu$. 
We develop the model in which the SM is coupled to scale invariant sector  in terms of DP with spin 1. We study the effects on the electroweak sector from the conformal sector, and the new bounds on DM physics are predicted. The coupling to the scale invariant degrees of freedom (d.o.f.) - the scalar dilaton sector - is most important operator in our analysis.  

In the classical scale invariant scheme at very high energies the dilaton $\bar \sigma$ appears in the gravity action (see, e.g., [2] and the footnote 1 thereby)
$$P = \int d^{4} x \sqrt {-\tilde g}\,\frac{1}{2}\left [\kappa r^{2} + (\partial_{\mu} \bar\sigma)^{2} - \eta\bar\sigma^{2}\,r\right ] $$
with the curvature $r$ and positive dimensionless parameters $\kappa$ and $\eta$. 
Once the conformal invariance is broken, the dilaton gets the vacuum expectation value (vev) $f$
and the last term in $P$ yields the Einstein-Hilbert action at low energies. The breaking of conformal invariance at the scale $\Lambda_{CFT} = 4\,\pi\,f$ triggers electroweak symmetry breaking (EWSB) at the scale $\Lambda_{EW} = 4\,\pi\,v < \Lambda_{CFT}$, where $v =246$ GeV is the vev of the Higgs boson. The scales $f$ and $v$ are different except for the Higgs boson ($f = v$).
The dilaton operator triggers the breaking of  $SU(2)_{L}\times U(1)_{Y}$ gauge invarince through  the dilaton mass operator. 

The hidden DM sector  can be explored in collider experiments. The dark photon mixes with the standard photon via the kinetic term $\sim \epsilon\,F_{\mu\nu}\,B^{\mu\nu}$ with $\epsilon$ being the  mixing strength; $F_{\mu\nu}$ and $B_{\mu\nu}$ are the strength tensors of electromagnetic $A_{\mu}$ and  $B_{\mu}$ fields, respectively. The basic object is
 the decay  $H\rightarrow \gamma\,\gamma^{\star}$, where $H$ should either be the SM Higgs boson $H$ or the scalar dilaton of conformal sector.


In this paper, we consider the scenario, where DM particle is the only DP within the reach of the LHC energy $\sqrt s \sim O(10~TeV)$. Recently, the DP and resonant monophoton signatures in Higgs boson decays at the LHC has been studied in [3].
The combined overall signal strength for the events of the Higgs boson decaying to two photons and four leptons 
obtained at the LHC by the ATLAS and CMS collaborations is $\mu  = 1.24 + 0.18 - 0.16$ [4] which may indicate a slight excess in the signal yields relative to the SM predictions. 
From the first glance, the slight deviation from $\mu =1$ (SM)  might be explained by the additional contribution of New Physics  effects, e.g., by quarks of 4th generation or by new scalar objects (e.g., charged Higgs bosons in 2-Higgs Doublet Model (2HDM)) in the loop. However, the extra quark generation is excluded by present Higgs data [5], while 2HDM has to be confirmed experimentally first. 

We consider the theory with the conformal anomaly of the type scalar-vector-vector that involves a scale (dilatation) current $K_{\mu}$ and two vector states $\gamma$ and $\gamma^{\star}$. The latter state should carry the scale invariant d.o.f.  In QCD the conformal anomaly reflects the violation of conformal invariance because of quantum effects: the divergence of $K_{\mu}$ does not vanish that indicates the breaking of scale invariance
$$\partial_{\mu}K^{\mu} = \theta^{\mu}_{\mu} = \frac{\beta (g)}{2\,g}\, G^{a}_{\mu\nu}\,G^{\mu\nu\,a} + \sum_{q} m_{q} [1 + \gamma_{m} (g^{2}) ]\bar{q}\,q.$$
Here, $\theta^{\mu}_{\mu}$ is the trace of the energy-momentum tensor, $\beta (g)$ is the standard QCD beta-function with the coupling constant $g$, $G^{a}_{\mu\nu}$ is the strength of gluon tensor. The quark state $q$ is accompanied by the mass $m_{q}$ and $\gamma_{m}(g^{2})$ stands for the anomalous dimension of the mass operator $\bar q q$. If the conformal invariance is breaking, the scalar color-singlet state $o (p)$ with the mass $m_{o}$,  the momentum $p_{\mu}$ and the decay constant $f_{0}$ can be produced when $K_{\mu}$ acts on the vacuum:
\begin{equation}
\label{e01}
 \langle 0\vert K^{\mu} (x)\vert o(p)\rangle = i\,p^{\mu}\,f_{o}\,e^{-i\,p\,x},\,\,\, \langle 0\vert \theta^{\mu}_{\mu}(x)\vert o(p)\rangle = m^{2}_{o}\,f_{o}\,e^{-i\,p\,x}. 
\end{equation}
The state $o$ in (\ref{e01})  would be either $\bar\sigma$ or the SM Higgs boson $H$ at low energies; $\vert 0\rangle$ is the vacuum state corresponding to spontaneously broken dilatation symmetry.
The coupling of the triangle QCD scale anomaly to $\gamma\,\gamma^{\star}$ final state is described by the effective interaction
$$L_{H\,\gamma\,\gamma^{\star}} = \epsilon\,g_{H\,\gamma\,\gamma^{\star}}\,F_{\mu\nu}\,B^{\mu\nu}\,H, $$
where the origin of $H$ is undestood through $\theta^{\mu}_{\mu}$ due to conformal anomaly; $g_{H\,\gamma\,\gamma^{\star}}$ is the coupling of the effective interaction 
related (proportional) to $f^{-1}_{o}$ and can be fixed from the experiment.

We suppose that an enhanced rate of the Higgs signal in $\gamma\gamma$ channel with respect to the SM prediction is due to scale invariant breaking sector of Conformal Field Theory (CFT), where the contribution from the DP with the mass $m$ to the branching ratio of the  Higgs boson decay $H\rightarrow\gamma\gamma$   would be significant [6]
$$BR(H\rightarrow\gamma\gamma^{\star})\simeq (1+ a\,\epsilon^{2}\,\Omega) BR^{SM}(H\rightarrow\gamma\gamma), $$
where $a \sim O(1)$ is the positive constant, $\Omega \sim (1 - m^{2}/m^{2}_{H})^{3}$, $m_{H}$ is the mass of $H$ and $BR^{SM}$ is the branching ratio of the decay $H\rightarrow\gamma\gamma$ within the SM expectations. The mass of the DP is unknown, however, the mixing factor $\epsilon$ is predicted in various models with the values in the range $10^{-12} - 10^{-2}$. In the low energy experiments the  values of $\epsilon$ in the area $10^{-5} - 10^{-2}$ have been probed [7]. On the phenomenological grounds, the DP masses in MeV range are favored for the present work. 
If no excess events are found, the obtained results would give the bounds on $\epsilon$ as a function of $m$.

The salient feature of the decay $H\rightarrow\gamma\gamma^{\star}$ is that the energy of $\gamma^{\star}$  has a continuous spectrum in the rest frame of $H$, in contrast to  $H\rightarrow\gamma\gamma$ in the SM. Therefore, by measuring the photon energy spectrum in the Higgs boson decay, one can discriminate the presence of DP or not. Signals of DP can be detected in the missing energy and momentum distribution carried away by DP once it was produced in  $H\rightarrow\gamma\gamma^{\star}$. Due to CP invariance the spin-1 nature of the missing energy is confirmed.
The interference effects with the amplitude of $H\rightarrow\gamma\gamma$ can also be taken into account. Finally, once DP is produced, the energy spectrum of the photon is no longer a $\delta$-like function peaked at $m_{H}/2$ as in   $H\rightarrow\gamma\gamma$, but rather a continuous one spreading from zero to $m_{H}/2$. 

In this paper DP is considered in the framework of the gauge dipole field model which exhibits an infrared (IR) singularities. In Abelian Higgs model the breaking of the gauge symmetry implies the dipole singularity in two-point Wightman function (TPWF) of the dipole fields, e.g., the scalar  or the gauge fields,  satisfying the equations of motion of 4th order [8-13]. The interacting dipole fields are local, relativistic quantum fields with a genuinely indefinite metric on the space of states generated from the vacuum. 
They converge asymptotically to free dipole fields. 
One of the crucial points of dipole quantum fields is that the massless dipole fields in $d=4$ space-time have a logarithmic increase at spatially separated arguments that features the confinement-like phenomenon [12,13]. Other classes of models exhibiting  IR singularities of the type $\delta^{\prime} (p^{2})$ in $p$-momentum space  have been presented in [14,15].

In Sec. II the couplings and constraints of the DM through the observable $\epsilon$ are found. Sec. III is devoted to the Lagrangian which defines the Higgs-dilaton Abelian gauge model, and to the resulting equations of motion. In Sec. IV we obtain TPWF and the propagator of DP. In the concluding section the results are summarized.

{\bf II. Couplings and constraints.-} CFT can be coupled to SM, and will then be a candidate to describe the DM sector. The ultraviolet (UV) coupling of an operator $O_{UV}$ of dimension $d_{UV}$ to a SM operator $O_{SM}$ of dimension $d$ at the UV scale $M$ (UV messenger) has the form
\begin{equation}
\label{e2}
\frac{1}{M^{d-4}} \,O_{SM} \frac{1}{M^{d_{UV}}}\,O_{UV}.
\end{equation}
No masses are allowed in the Lagrangian of the effective theory containing (\ref{e2}). All masses  can be generated dynamically in IR.
We consider the hidden sector which is formed itself when the dilaton field $\bar\sigma$ is coupled to a $U(1)$ gauge theory. The coupling of $\bar\sigma (x)$ to DM sector in UV is
$$\frac{1}{M^{d_{UV}-2}} \,{\vert\bar\sigma\vert }^{2}\,O_{UV},$$
which flows in IR to coupling of the Higgs boson $H$ to DM operator $O_{IR}$ of dimension $d_{IR}$ 
$$const\,\frac{\Lambda^{d_{UV} - d_{IR}}}{M^{d_{UV}-2}} \,{\vert H\vert }^{2}\,O_{IR}$$
when the scale invariance is almost breaking ($\Lambda$ is the strong coupling scale). Once $H$ acquires $v$, theory becomes nonconformal below the scale $\tilde\Lambda$, where [16]
$$\tilde\Lambda^{4- d_{IR}} = \frac{\Lambda^{d_{UV} - d_{IR}}}{M^{d_{UV}-2}} \,v^{2}.$$
Below $\tilde\Lambda$ the DM sector becomes a standard particle sector. For a typical energy $\sqrt {s}$ of a collider experiment $\tilde\Lambda < \sqrt {s} <\Lambda$, which leads to the energy constraint  to search for DM physics
$$s^{2- d_{IR}/2} > \left (\frac{\Lambda}{M}\right )^{d_{UV} - d_{IR}} \,M^{2 - d_{IR}} \,v^{2}.$$
Based on the operator form $(\ref{e2})$ the mixing strength $\epsilon$, as an observable, is 
\begin{equation}
\label{e8}
\epsilon = \left (\frac {\sqrt {s}}{M}\right )^{2(d-4)}\, \left (\frac {\sqrt {s}}{\Lambda}\right )^{2d_{IR}}\, \left (\frac {\Lambda}{M}\right )^{2d_{UV}}. 
\end{equation}
Then the effect of DM sector on observables has no the dependence on $d_{UV}$, $d_{IR}$ and $\Lambda$, and is bounded by [6]
\begin{equation}
\label{e9}
\epsilon < \frac{s^{d}}{\left (v^{2}\,M^{d-2}\right )^{2}}.
\end{equation}
It is clear from (\ref {e8}) that signals of new physics with DM increase with energy, the dimension $d$ of the SM operator, and would be seen if the values of the parameter $M$ are not too large. If the deviation from the SM is detected at the level of order $3\%$, the DM would be visible at the LHC ($\sqrt{s}\sim$ O(10 TeV)) as long as $M < 1000$ TeV with  $d=4$.
The result for an upper limit (\ref{e9}) is shown in Fig.1 for $\sqrt {s}$ = 8 -14 TeV and $M$ = 800-1000 TeV ($d=4$). 
\begin{figure}[h!]
\renewcommand{\figurename}{Fig.}
\centering
 \includegraphics[width=\textwidth]{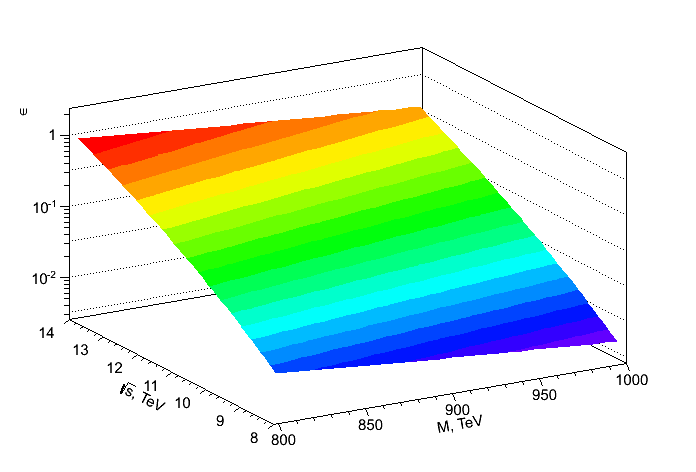}
 \caption{{ \it Upper limit of $\epsilon$ (\ref{e9}) as a function of $\sqrt {s}$ and $M$. }}
 \label{fig:secondgraph}
\end{figure}


Assuming that the production cross section of $H$ is dominated by the gluon fusion process, we propose the measured value of $\mu$ as 
$\mu = (1+ a\,\epsilon^{2}\,\Omega) F$, where $ F = c^{2}_{G}\,v^{2}/f^{2}$ is the ratio between the significances of the dilaton and the SM Higgs boson signals; $c_{G} = 11 - 2\,n_{light}/3$, $n_{light}$ is the number of quarks (in the loop) lighter than the dilaton. The upper limit of the DP mass can be estimated through the formula
\begin{equation}
\label{e88}
m^{2} < m^{2}_{H}\left \{ 1 - \left [\frac{1}{a\,x}\left ( \frac{\mu}{F} -1 \right )\right ]^{1/3}\right \}, 
\end{equation}
where $ x$ is an upper limit of the mixing strength squared. 
In the decay $H\rightarrow\gamma\gamma^{\star}$ the gauge invariant operator structure is
 $O_{SM}\,O_{IR}\sim \epsilon\bar q\,\gamma_{\mu}\,q\,H\, B^{\mu}\,M^{-1}$
and the relevant energy scale  is the mass of the heavy quark in the loop. 
Since the new effects beyond the SM is expected on the scale $M > v \sim O(0.3 ~ TeV)$, we find the upper limit on the mixing strength $\epsilon < 3\cdot 10^{-2}$
in the case of the top-quark in the loop  ($d=4$).
Once the mixing $\epsilon$ is almost zero, the only decay of the SM Higgs boson into two photons would be appropriate. 

The amplitude of the decay $\gamma^{\star}$ into neutrino-antineutrino pair, $\nu\bar\nu $, is
$$Am (\gamma^{\star}\rightarrow \nu\bar\nu) = \frac{1}{2}\,f_{\nu}\,\bar\nu \left (g_{V_{\nu}}\,\gamma_{\beta} + g_{A_{\nu}}\,\gamma_{\beta}\,\gamma_{5} \right)\,\nu\,\gamma^{\star}_{\beta}, $$
where  $f^{2}_{\nu} = 4\,\sqrt {2} \,G\,m^{2}$, $G\sim 10 ^{-5}~GeV^{-2}$ is the weak coupling strength.
Since there are no final state interactions in $\gamma^{\star}\rightarrow \nu\bar\nu $ decay, one has its partial width in the form 
$\Gamma (\gamma^{\star}\rightarrow \nu\bar\nu) =(2/3)\,\bar\alpha\,\epsilon^{2}\cdot m$, where 
$\epsilon^{2} =  {\sqrt{2}\,G\,g^{2}_{\nu}\,m^{2}}/{(4\,\pi\,\bar\alpha)}$, $\bar\alpha$ is an electromagnetic gauge coupling, 
and $g^{2}_{V_{\nu}} = g^{2}_{A_{\nu}} \equiv g^{2}_{\nu}$ is taken into account. 
Using the  restriction on $\epsilon < 3\cdot 10^{-2}$ above mentioned, one can find the upper limit on the DP mass $m < 4.7 $ GeV.  There is a weak dependence on $f$ from the DP mass $m$. The main contribution is given by the increasing factor $c_{G}\sim 7$ for either $n_{light} = 5$ or $n_{light} = 6$.
The combined data of ATLAS and CMS [4] with the central value  $\mu = 1.24$ gives the maximal value of  $f$  on the level of 1.54 TeV (see (\ref{e88})) in the case when all the known quark flavors are lighter than the dilaton.
If the latter is lighter than the top-quark, we conclude $f < 1.7$ TeV.  Comparing the result for $\Gamma (\gamma^{\star}\rightarrow\nu\bar\nu)$ above mentioned and the calculation of the process $\gamma^{\star}\rightarrow\nu\bar\nu$ with electromagnetic neutrino formfactor (EMNF) [17], we find the DP mass in the form
$$ m = m_{l} \left [ \frac{18\,\pi^{2}}{\bar\alpha^{2}\,\left (\ln \frac{\Lambda^{2}_{\nu}}{m_{l}^{2}} -\frac{1}{6}\right)^{2}}\right ]^{\frac{1}{4}}.$$
Here, $m_{l}$ is the mass of the charged lepton in the loop, $\Lambda_{\nu}$ is the cut off scale in the logarithmically divergent integral of EMNF. Using $\Lambda_{\nu}$ at the scale of the $Z$-boson mass, the DP mass $m = 4.5$ MeV is equated for the case of an electron loop in EMNF.
 The LHC is the good facility where the DM physics can be tested well.

{\bf III. Model.-} The  Lagrangian density (LD) of  the Higgs-dilaton Abelian gauge model  where the  field $B_{\mu}$ mixes with  $A_{\mu}$ is
\begin{equation}
\label{e10}
L_{\epsilon} = -\frac{1}{2}\,\epsilon\,F_{\mu\nu}\,B^{\mu\nu} - \xi (\partial_{\mu} A^{\mu})(\partial_{\nu} B^{\nu}) + \bar q (i\hat\partial - m_{q} - g\,\hat A)\,q - I^{\mu} (B_{\mu} - \partial_{\mu}\sigma),
\end{equation}
where 
$\xi$ is a real number, $g$ is dimensioneless coupling constant,  
$F_{\mu\nu} = \partial_{\mu} A_{\nu} - \partial_{\nu} A_{\mu}$, 
$B_{\mu\nu} = \partial_{\mu} B_{\nu} - \partial_{\nu} B_{\mu}$, 
 and $I_{\mu}$ is an auxiliary field.  
The subcanonical scalar field $\sigma (x)$ with zero dimension in mass units is the primary dilaton field (the grandfather  potential) which provides a control over UV and IR divergences.
As the true Nambu-Goldstone boson, the dilaton couples to other fields via the derivative, thus avoiding bounds on a fifth force. LD (\ref{e10}) is invariant under the restricted gauge transformations of the second kind
\begin{equation}
\label{e11}
A_{\mu}\rightarrow A_{\mu} +\partial_{\mu}\alpha, \,\,\, B_{\mu}\rightarrow B_{\mu} +\partial_{\mu}\alpha,\,\, \sigma\rightarrow \sigma + \alpha,\,\, \psi\rightarrow \psi\,e^{i\,g\,\alpha},\,\, I_{\mu}\rightarrow I_{\mu}, 
\end{equation}
where $\alpha (x)$ obeys the equation $\Box\alpha (x) = 0$ ($ \Box \equiv \partial_{\mu} \,\partial^{\mu}$).  
The parametrization of the $H$ couplings to quarks that are relevant for collider physics is
\begin{equation}
\label{e12}
L_{H} = -\frac{H}{v}\,\sum_{q} m_{q} \,q\,\bar q.
\end{equation}
If the SM is the part of CFT the Higgs couplings to massless gauge bosons with $q$-fields in the loop is replaced by corresponding dilaton couplings  $(2\,m^{2}_{q}/v^{2})H^{+}\,H (x) \rightarrow m^{2}_{q}\,\sigma^{2}(x)$ [18].
Hence, (\ref{e12}) is transformed to 
\begin{equation}
\label{e122}
L_{\sigma} = -\frac{\sigma}{\sqrt{2}}\sum_{q^{\prime}} (m_{q^{\prime}} + x_{\sigma}\,y_{q^{\prime}}\,v)\,q^{\prime}\,\bar q^{\prime},
\end{equation}
where  $x_{\sigma} = m^{2}_{\sigma} /f^{2} <1$ parametrizes the size of deviations from exact scale  invariance; $m_{\sigma}$ is the mass of the dilaton; $y_{q}$ are 9 additional contributions to the Yukawa couplings. If SM is embedded in the conformal sector we assume that  $q^{\prime}$ are those quark d.o.f. which obey the condition 
\begin{equation}
\label{e1222}
\sum_{light} b_{i} + \sum_{heavy} b_{i} = 0, 
\end{equation}
where $i$ carries either QCD or electroweak (EW) features of the coefficients $b_{i}$ of corresponding $\beta$-functions.
The sum in (\ref{e1222}) is splitted over all colored particles into sums over light and heavy states in the mass scale separated by  $m_{\sigma}$. 
Hence the only quarks $q^{\prime}$ in (\ref{e122}) lighter than that of the dilaton are included in the coefficients of corresponding $\beta$-function. 
In particlular, $b^{light}_{QCD} = - c_{G}$, where the number of light quarks is either  $n_{light} = 5$ if $m_{\sigma} < m_{t}$, or $n_{light} = 6$ if $m_{\sigma} > m_{t}$ for the top quark mass $m_{t}$ [18]. In EW sector one has $b^{light}_{EW} = -80/9$ if $m_{\sigma} < 2 m_{W}$, or $b^{light}_{EW} = -35/9$ if $2m_{W} < m_{\sigma} < 2 m_{t}$, or $b^{light}_{EW} = -17/3$ if $2 m_{t} < m_{\sigma}$, where $m_{W}$ is the mass of the $W$-boson [19].
Once the conformal invariance is breaking down, the dilaton is the Higgs-boson itself when $f=v$. The dilaton $\sigma$ serves as  a conformal compensator which under gauge transformation shifts as a Goldstone boson, $\sigma\rightarrow \sigma + \alpha$.
The invariance of LD (\ref{e10}) is broken because of (\ref{e12}), (\ref{e122}) and a small mass parameter $m$ of DP if they should incorporate in (\ref{e10}).  Finally, because of conformal condition (\ref{e1222}) the dilaton contribution to the decay $H\rightarrow\gamma\,\gamma$ corresponds to including the states lighter than the dilaton in the $\beta$-function coefficients.

Consider the scale invariant sector of the theory, which contains the  fields $H$ and $\bar\sigma$ in the action
\begin{equation}
\label{e13}
P(H,\bar\sigma) =\int d^{4} x L_{SI} (x) = \int d^{4} x\left \{\frac{1}{2}\left [\left (\partial_{\mu} H\right )^{2}  + \left (\partial_{\mu} \bar\sigma\right )^{2}\right ] - \frac{\lambda}{4}\left (H^2 -\beta^2 \bar\sigma^{2}\right )^{2}\right\}.  
\end{equation}
The action (\ref{e13}) is invariant under dilatation transformations $x\rightarrow x^{\prime} = c\,x$ $(c > 0)$, $\phi (x) \rightarrow \phi^{\prime} (x^{\prime}) = c^{-d_{\phi}}\,\phi (x)$, where $d_{\phi}$ is the scaling dimension of $\phi$, where $\phi: H,\,\bar\sigma$;  $\beta = \langle H\rangle /\langle\bar\sigma\rangle = v/f$. 
The LD of the model becomes
\begin{equation}
\label{e14}
L = L_{\epsilon} + L_{\sigma} -\frac{1}{2}\,m ^{2}\,B_{\mu}\,B^{\mu} + L_{SI}. 
\end{equation}

The equations of motion are
$$(i\hat\partial - m_{q} -
g\hat A)\,q =0,\,\,\,   \sigma\sum_{q^{\prime}} (m_{q^{\prime}}  + x_{\sigma}\,y_{q^{\prime}}\,v)\,q^{\prime} = 0,$$
$$\partial_{\mu}\,I^{\mu} = - f\,\beta^{2}\,\frac{\bar\sigma}{H}\,\Box H - f\,\Box\bar\sigma - \sum_{q^{\prime}} (m_{q^{\prime}} + x_{\sigma}\,y_{q^{\prime}}\,v)\bar q^{\prime}\, q^{\prime},$$
\begin{equation}
\label{e17}
\Box B_{\mu} = \frac{1}{\epsilon} Y_{\mu} + \left (1-\frac{\xi}{\epsilon}\right ) \partial_{\mu}(\partial\cdot B),\,\,\, Y_{\mu} =g\bar q\,\gamma_{\mu}\,q,
\end{equation}
\begin{equation}
\label{e18}
B_{\mu} = \frac{\epsilon}{m^{2}} \left [\Box A_{\mu} - \frac{1}{\epsilon}\,I_{\mu} -\left (1-\frac{\xi}{\epsilon}\right ) \partial_{\mu}(\partial\cdot A)\right ].
\end{equation}
The equation of motion for $q$ implies current conservation $\partial_{\mu} Y^{\mu} = 0$. 
Then taking into account the equations  (\ref{e17}) and  $B_{\mu}(x) = \partial_{\mu}\sigma (x)$, one approaches the dipole equation for the Hermitian virtual dilaton  field $\bar\sigma (x)$  
\begin{equation}
\label{e21}
\lim_{x_{\sigma}\rightarrow 0} \Box^{2}\bar\sigma (x) \simeq 0,\,\,\,\, f\neq 0.
\end{equation}


{\bf IV. TPWF and propagator.-} It is known that the space with an indefinite metric is useful to the formalization of the idea of virtual (or potential) states. For this, the Hilbert space of physical state vectors has to be replaced by pseudo-Hilbert space of virtual state vectors.
The TPWF for $\bar\sigma (x)$ has the form
\begin{equation}
\label{e22}
\omega (x) = \langle\Omega, \bar\sigma (x)\, \bar\sigma (0)\Omega\rangle.
\end{equation}
 The vacuum vector $\Omega$ in (\ref{e22}) is defined as the vector satisfying $\bar\sigma^{(-)} (x)\Omega = 0$, $\langle\Omega,\Omega\rangle = 1$, where $\bar\sigma (x)$ is decomposed into negative-frequency (annihilation) and positive-frequency (creation) parts: $\bar\sigma (x) = \bar\sigma^{(-)} (x) + \bar\sigma^{(+)} (x)$, $\bar\sigma^{(+)} (x) = [\bar\sigma^{(-)} (x)]^{+}$.  
The function $\omega (x)$ (\ref{e22}) should be Lorentz-invariant and the equation 
\begin{equation}
\label{e23}
\Box^{2} \omega (x) = 0
\end{equation}
is evident from (\ref{e21}). The general solution of (\ref{e23}) is the following expansion (see [10] and the refs. therein)
\begin{equation}
\label{e24}
\omega (x) = b_{1}\,\,ln\frac{l^{2}}{-x^{2}_{\mu} +i\,\varepsilon\,x^{0}} + b_{2}\,\frac{1}{x^{2}_{\mu} - i\,\varepsilon\,x^{0}} + b_{3}
\end{equation}
which is the distribution on the space $S^{\prime}(\Re ^{4})$ of temperate generalized function on $\Re ^{4}$. The coefficients $b_{1}$ and  $b_{2}$ in (\ref{e24}) will be defined later, while $b_{3}$ is an arbitrary constant.
The parameter $l$ in (\ref{e24}) having the dimension in units of length  breaks the scale invariance under dilatation transformation that implies spontaneously symmetry breaking of the dilatation invariance of (\ref{e21}). 
The TPWF (\ref{e22}) is the homogeneous generalized function of the zeroth order with the dilatation properties 
$\omega (\rho x) = \omega (x) - 1/(8\,\pi)^{2}\,\ln\rho,\,\,\,\rho > 0. $

The Fourier transformation (FT) of the first term in (\ref{e24}) is proportional to $\int 2\,\pi\,\theta (p^{0})\,\delta^{\prime}(p^{2}, \hat M^{2})\,e^ {-ipx} d_{4}p$ which does not have the (positive) measure (IR divergence), where $\hat M = (2/l)\, e^ {1/2-\gamma}$, $\gamma $ is the Euler's constant. 
The $\delta^{\prime} (p^{2}, \hat M^{2})$ is well-defined distribution only on the space $S(\Re ^{4})$ of complex Schwartz test functions on $\Re ^{4}$
$$ \delta^{\prime}(p^{2}, \hat M^{2}) = \frac{1}{16} \Box_{p}^{2} \left [ \theta (p^{2})\,\ln \frac{p^{2}}{\hat M^{2}}\right ], $$
and would have the form as $\delta^{\prime} (p^{2})$ for the same IR reasons only on the space
$S_{0}(\Re ^{4})$, where the test functions $u(x)$ from $S(\Re ^{4})$ are zero in the origin, $u(0) = 0$. The distribution  $\delta^{\prime} (p^{2})$ is defined by 
$$\int 2\,\pi\,\theta (p^{0})\,\delta^{\prime}(p^{2})\,u (p)\, d_{4} p = \int_{\Gamma_{0}^{+}} \, \frac{1}{2\,n\,p} \,\left ( -n\,\partial + \frac{1}{n\,p}\right ) \,u (p) \, (d p)_{0}, $$
where $n$ is the fixed unit time-like vector ($n^{2} =1 $) in the Minkovsky space $\mathcal {S} (\mathbb {M})$ from $V^{+} = \{p \in\mathbb {M}, p^{2} > 0, p^{0} > 0\}$; $\Gamma_{0}^{+} = \{p \in \mathbb {M}, p^{2} = 0, p^{0} > 0\}$; $u(p)$ is an arbitrary function from $ \mathcal {S} (\mathbb {M})$ which is zero at $p = 0$. For $n = (1, \vec {0})$ one has
$$- \frac{\partial}{\partial m_{\sigma}^{2}} \int 2\,\pi\,\theta (p^{0})\,\delta(p^{2} - m_{\sigma}^{2})\,u (p)\, d_{4} p = - \frac{\partial}{\partial m_{\sigma}^{2}} \int \frac{d_{3} p}{2\,E} \,u (E, \vec {p}). $$
One can verified that
$(p^{2})^{2}\,\delta^{\prime} (p^{2}, \hat M^{2}) = 0, \,\,\, p^{2}\,\delta^{\prime} (p^{2}, \hat M^{2}) = - \delta (p^{2}).$
The presence of $\delta^{\prime} (p^{2}, \hat M^{2})$ in FT of TPWF (\ref{e24}) is a consequence of the nonunitarity of translations ($\delta^{\prime} (p^{2}, \hat M^{2})$ is not a measure). 

The commutator for dilaton field  is
\begin{equation}
\label{e25}
[\bar\sigma (x), \bar\sigma (0) ] = 2\,\pi\,i\,sign (x^{0}) \left [b_{1} \,\theta (x^{2}) + b_{2}\,\delta (x^{2})\right ]. 
\end{equation}
The coefficients $b_{1}$ and $b_{2}$ in (\ref{e24}) and (\ref{e25}) can be fixed from the canonical commutation relations 
\begin{equation}
\label{e26}
 \left [A_{\mu}(x), \pi_{A_{\nu}} (0)\right ]_{\vert x^{0} = 0} = i\,g_{\mu\nu}\,\delta^{3}(\vec x)
\end{equation}
and 
\begin{equation}
\label{e27}
 \left [\bar\sigma (x), \pi_{\bar\sigma} (0)\right ]_{\vert x^{0} = 0} = i\,\delta^{3}(\vec x),
\end{equation}
respectively. $\pi_{\bar\sigma} (x)$ and $\pi_{A_{\mu}} (x)$ in (\ref{e27}) and (\ref{e26}) are the conjugate momenta of $\bar\sigma (x)$ and $A_{\mu} (x)$, respectively.
Thus, the definition of the dilaton field $\bar\sigma (x)$ by means of the secondary quantization method must be performed in the space with an indefinite metric.

In order to find the coefficients $b_{1}$ and $b_{2}$ we choose  $I_{\mu}$ in the form 
$I_{\mu} = \epsilon\,m^{2} (\partial_{\mu}\sigma - A_{\mu})$ which is invariant under  gauge transformations (\ref{e11}). Hence, we have an additional mixing term in LD (\ref{e14}), namely $ \epsilon\,m^{2} (A_{\mu} - \partial_{\mu}\sigma) (B_{\mu} - \partial_{\mu}\sigma )$ accompanied by the term 
$\sim \epsilon\,F_{\mu\nu}\,B^{\mu\nu}$. The new equations arise
\begin{equation}
\label{e28}
\Box\,A_{\mu} - a\,\partial_{\mu}\,(\partial A) +m^{2} (A_{\mu} - \epsilon^{-1}\,B_{\mu}) = m^{2}\,\partial_{\mu}\sigma,\,\,\, a = 1 - \xi/\epsilon,
\end{equation}
\begin{equation}
\label{e29}
(\Box + \tilde m^{2} )(\partial B) = \tilde m^{2} \Box\sigma,\,\,\, \tilde m^{2} = \frac{\epsilon}{\xi}\,m^{2} 
\end{equation}
instead of Eqs. (\ref{e18}) and (\ref{e17}), respectively. The solution $B_{\mu} = \partial_{\mu}\sigma$  obtained for an arbitrary vector $I_{\mu}$ obeys Eq. (\ref{e29}) which leads to (\ref{e21}).
The solution of Eq. (\ref{e28}) is explicitly given in the form 
\begin{equation}
\label{e30}
 A_{\mu} = C_{\mu} + \frac{1}{m}\,\partial_{\mu}\varphi  - \frac{\xi}{\epsilon\,m^{3}}\,\partial_{\mu}\Box\varphi,
\end{equation}
where $\varphi = (1 + \epsilon^{-1})\,m\,\sigma$ with $(\Box +m^{2})C_{\mu} = 0$, $\partial_{\mu} C^{\mu} = 0$ and $[C_{\mu} (x),\varphi (y) ] = 0$. 

Using (\ref{e30}) in (\ref{e26}) with  $\pi_{A_{\nu}} = \epsilon (\partial_{0} B_{\nu} - \partial_{\nu} B_{0} )- \xi\,g_{0\nu}\,\partial^{\rho} B_{\rho}$ one can find 
 $$b_{1} = \frac{1}{(2\,\pi)^{2}}\,\frac{\epsilon\,f^{2}}{\xi\, (1 + \epsilon)}$$
while (\ref{e27}) with 
$$\pi_{\bar\sigma} = \left (1 + \frac{\epsilon\,m^{2}}{f^{2}}\right)\,\partial_{0}\bar\sigma - \frac{\epsilon\,m^{2}}{f}\,A_{0} $$ gives 
 $$b_{2} = \frac{-1}{2\,\pi^{2}\,\left (1 - m^{2}/f^{2}\right )}.$$
The propagator of canonical grandfather potential $\bar\sigma (x)$ is
\begin{equation}
\label{e33}
 \tau (x) =  \langle\Omega, T [\bar\sigma (x) \bar\sigma (0)] \Omega\rangle 
  = - b_{1}\left [\ln\vert \kappa^{2} x^{2}\vert + i\pi\theta (x^{2})\right ] +b_{2} 
\left [\frac {1}{x^{2}} + i\pi\delta (x^{2})\right ] + b_{3}, 
\end{equation}
where $\kappa \sim l^{-1}$. On the other hand, (\ref{e33}) can be obtained through the Fourier transformed distribution 
\begin{equation}
\label{e34}
\tau (x)  = \frac{(2\pi)^{2}}{i} \int d_{4}p e^{-ipx}\left \{4b_{1}\lim_{\lambda^{2}\rightarrow 0}\left [\frac{1}{(p^{2} -\lambda^{2} + i\varepsilon)^{2}} +i\pi^{2}\ln\frac{\lambda^{2}}{\kappa^{2}}\,\delta_{4}(p)\right ] + \frac{b_{2}}{p^{2} + i\varepsilon}\right \}.
\end{equation}
To get (\ref{e34}) we have used the following properties of generalized functions [20]
$$\int d_{4}p\,e^{-i\,p\,x} \,\frac{1}{(p^{2} - \lambda^{2} + i\,\varepsilon)^{2}} = \frac{i}{8\,\pi^{2}} K_{0}\left (\lambda\,\sqrt{-x^{2}_{\mu} + i\,\varepsilon}\right ) $$
for small argument of the Bessel function $K_{0}(z)$
\begin{equation}
\label{e36}
K_{0}(z)\simeq \ln\left (\frac{2}{z}\right) - \gamma + O(z^{2}, z^{2}\,\ln z).
\end{equation}
The coefficient $b_{3}$ in (\ref{e33}) is fixed in such a way as to cancel the term proportional to $\ln 2 - \gamma$ in (\ref{e36}). The second term in the propagator (\ref{e34}) proportional to $(1 + m^{2}/f^{2}) (p^{2} + i\,\varepsilon)^{-1}$ is not a distribution (generalized function) since it is IR divergent. This term also breaks the UV stability that means
$$ (-p^{2})^{2}\,\lim_{\lambda^{2}\rightarrow 0} \left [\frac{1}{(p^{2} -\lambda^{2} + i\varepsilon)^{2}} +i\pi^{2}\ln\frac{\lambda^{2}}{\kappa^{2}}\,\delta_{4}(p)\right ] = 1,$$
$$ \frac{1}{(4\,\pi)^{2}}\,\Box^{2} \, \left \{\ln\vert \kappa^{2}\,x_{\mu}^{2}\vert + i\,\pi\,\theta (x^{2}) \right \} = \delta^{4} (p).$$

The commutator of $B_{\mu}$ field is 
$[B_{\mu} (x), B_{\nu} (y) ] = \left (1 + 1/\epsilon\right )^{-2}\,[A_{\mu} (x), A_{\nu} (y)],$
where 
$[A_{\mu} (x), A_{\nu} (y)] = -2\,i/(\pi\,\xi)\,\left (1 + 1/\epsilon\right )\,g_{\mu\nu} \,\varepsilon (x^{0})\,\delta (x^{2}).$

The TPWF $\omega^{B}_{\mu\nu} (x)$ for $B_{\mu}$ field is defined by TPWF  $\omega^{A}_{\mu\nu} (x)$ for $A_{\mu}$ field :
$\omega^{B}_{\mu\nu} (x) = - 4\,\epsilon/[\xi\,(1 + \epsilon)]\,\omega^{A}_{\mu\nu} (x),$
where
$$\omega^{A}_{\mu\nu} (x) = 2\,\pi\,\int d_{4} p\,e^{-i\,p\,x}\,\theta (p^{0}) \left [-g_{\mu\nu}\,\delta (p^{2}) + (\xi -1)\,\delta^{\prime} (p^{2})\,p_{\mu}\,p_{\nu}\right ]$$
in the arbitrary $\xi$ - gauge.
The propagator of $B_{\mu}$ field in four-momentum space is 
\begin{equation}
\label{e41}
\tilde\tau_{\mu\nu} (p) = -i\,\left [ \tilde\tau_{1_{\mu\nu}} (p) + \tilde\tau_{2_{\mu\nu}} (p)\right ],
\end{equation}
with
$$\tilde\tau_{1_{\mu\nu}} (p) = \frac{4}{\xi\,(1 + \epsilon^{-1})}\,p_{\mu}\,p_{\nu}\,\lim_{\lambda^{2}\rightarrow 0} \left [\frac{1}{(p^{2} -\lambda^{2} +i\,\varepsilon)^{2}} + i\,\pi^{2}\,\ln (l^{2}\,\lambda^{2})\,\delta_{4}(p)\right ],$$
\begin{equation}
\label{e43}
\tilde\tau_{2_{\mu\nu}} (p) = -\frac{2}{f^{2} - m^{2}}\,\frac{p_{\mu}\,p_{\nu}}{p^{2}  +i\,\varepsilon},
\end{equation}
where the "strong gauge condition"  $\partial_{\mu}\sigma (x) = B_{\mu} (x)$ was used when the FT of $\tau (x)$ has been evalulated in (\ref{e34}) first. 

Following by Zwanziger [11], the first term in (\ref{e41}) is understood through the generalized function with the weak derivative $\partial/\partial p^{\rho}$ in the integral
\begin{equation}
\label{e44}
\lim_{\lambda^{2}\rightarrow 0} \,\int d^{4} p \,\frac{\ln (p^{2}/\hat M^{2} - i\,\varepsilon)}{(-p^{2} + \lambda^{2} - i\,\varepsilon)^{2}}\,p^{\rho}\,\frac{\partial}{\partial p^{\rho}} \left [p^{\mu}\,p^{\nu}\,R_{\mu\nu} (p)\right ],
\end{equation}
where $R_{\mu\nu}(p)$ is  some test function. The operator $p^{\nu}\,\partial /\partial p_{\nu}$ in (\ref{e44}) annihilates all terms of degree zero in momentum $p$ which are  logarithmically divergent ones, whether in UV or in IR [11].
For small enough mixing $\epsilon$  the leading term in  (\ref{e41}) is given by (\ref{e43}).

{\bf V. Conclusions.-}  The model for the DM particle has been studied at the lowest order of perturbative gauge theory using canonical quantization. The DP solution has been found, moreover the vector DM field is massive. 
 The slight excess in the $\gamma\gamma$-signal of the Higgs boson relative to the SM estimation is explained by the influence of the conformal sector with DP contribution. The interaction between DP and quarks  is mediated by the derivative of the scalar, the dilaton.
The propagator of DP (\ref{e41}) turns to the standard form of massless photon once the mixing $\epsilon$ and the DP mass $m$ are disappeared. We estimated the upper limit for the mixing strength $\epsilon < 3 \cdot 10^{-2}$ for two-photon decay of the Higgs-like  boson where the main contribution is due to top-quark in the loop. This can be interpreted as the limit of the branching ratio  $BR (H\rightarrow\gamma\,\gamma^{\star})$ which is just the rate of the two-photon decay of the Higgs boson in the SM as $\epsilon = 0$. 
We find that the DP mass is restricted by 4.7 GeV from above. The theoretical approach based on the result $\Gamma (\gamma^{\star}\rightarrow\nu\bar\nu) = (2/3)\,\epsilon^{2}\,\bar\alpha\,m$ accompanied by  EMNF calculations gives for DP mass $m =4.5$ MeV. 
The combined data of ATLAS and CMS at LHC [4] can allow to report that $f < 1.7$ TeV for $n_{light} = 5$, while $f < 1.54$ TeV if the dilaton is heavier that the top quark. 
The decay mode $H\rightarrow\gamma\,\gamma^{\star}$ can be used to probe the DM sector since the emitted energy of the single photon is encoded with measuring of the missing of the recoil DP.
It's clear from (\ref{e8}) and (\ref{e9}) that the LHC is the most promising machine where DM physics can be discovered.

\end{document}